\documentstyle [11pt,leqno]{article} 
\begin{document}

\title{ORIENTING GRAPHS TO OPTIMIZE REACHABILITY\thanks{This work
was supported by the National Science Foundation under Grants
NCR-95-05551 and DMS-92-06991}}

\author{S. Louis Hakimi,\thanks{Department of Electrical and
Computer Engineering, University of California, Davis, CA 95616}
\and
Edward F. Schmeichel,\thanks{Department of Mathematics and Computer
Science, San Jose State University, San Jose, CA 95192}  
\and
Neal E. Young\thanks{Department of Computer Science, Dartmouth
College, Hanover, NH 03755}}

\date{ }


\maketitle


\begin{abstract} It is well known that every
2-edge-connected graph can be oriented so that the resulting
digraph is strongly connected. Here we study the problem of
orienting a connected graph with cut edges in order to maximize
the number of ordered vertex pairs (x,y) such that there is a
directed path from x to y. After transforming this problem, we
prove a key theorem about the transformed problem
that allows us to obtain a quadratic algorithm for the original
orientation problem. We also
consider how to orient graphs to
minimize the number of ordered vertex pairs joined by a directed
path. After showing this problem is equivalent to the comparability
graph
completion problem, we show both problems are NP-hard,
and even NP-hard to approximate
to within a factor of $1+\epsilon$, for some $\epsilon > 0$. 
\end{abstract} 

\setlength{\baselineskip}{14pt}
\section{\bf Introduction} 

\indent Our terminology and notation
is standard except as indicated. We
mention only that if X is a subset of the
vertices in a graph, we
use $\langle X\rangle$ to denote the
subgraph induced by $X$. Good references for any other undefined
terms are [1,2].

Let $G$ be any connected graph. Given an orientation $\vec{G}$ of
$G$, we will use $R(\vec{G})$ to denote the number of ordered
vertex pairs $(x,y)$ such that there is a directed path from $x$
to $y$ in $\vec{G}$. We call $R(\vec{G})$ the {\sl reachability}
of $\vec{G}$.

Robbins [12] showed that $G$ can be oriented so that $\vec{G}$
is
strongly connected (i.e., $R(\vec{G})=|V|(|V|-1))$ if and only if
$G$ is 2-edge-connected. In the following section (Section 2),
we study the problem of how to orient
an arbitrary graph $G$
to obtain a digraph $\vec{G}$ with $R(\vec{G})$ as
large as possible. We
first transform the problem into an
equivalent orientation problem on vertex-weighted trees. Although
this transformed version is NP-hard, we
prove a key theorem
(Theorem 2.2) which
allows us to get a quadratic algorithm
for the original problem, as well as a fully polynomial
approximation scheme for the transformed problem.

In Section 3, we consider the analogous problem of how to orient
$G$ so as to minimize $R(\vec{G})$ for the resulting digraph
$\vec{G}$. 
We show this problem is equivalent to comparability graph
completion
(adding the fewest edges so the resulting graph can be transitively
oriented), and then show that both problems are NP-hard, and even
NP-hard to approximate
within a factor of $1+\epsilon$, for some $\epsilon > 0$.
Related hardness results appear in [6,10].

\section{\bf Orienting Graphs to Maximize Reachability}

\indent Suppose we are given a connected graph $G$ on $n$
vertices with cut edges. Our goal is to orient $G$ to obtain a
digraph $\vec{G}$ with $R(\vec{G})$ as large as possible.

It will be  useful to first transform this basic problem. Let
$C_1, C_2,\linebreak \ldots, C_b$ denote the components left when
the cut edges of $G$ are removed from $G$. By Robbins Theorem
[12] each $C_i$ can be oriented into a strongly connected
digraph. Let us contract each non-trivial $C_i$ into a single
vertex $x_i$, giving $x_i$ weight $wt(x_i)=|V(C_i)|$. The
resulting contracted graph is, of course, a $b$-vertex tree
$T=T(G)$ with integer weights on the vertices. Our original
problem is now easily seen to be equivalent to the following
problem: How should we orient $T$ to maximize $\sum wt(x_i) \cdot
wt(x_j)$, the sum being taken over all vertex pairs $(x_i, x_j)$
in $T$ such that there is a directed path from $x_i$ to $x_j$ in
$\vec{T}$? Note that the input size of the transformed problem
may be exponentially smaller than the input size of the original
problem (roughly, $b(1+\mbox{log}\, n)$ versus $n$).

We begin by showing that this transformed problem is NP-hard. In
particular, consider the following decision problem:

\bigskip\noindent {\bf WEIGHTED TREE ORIENTATION (WTO)}
\begin{description}
\item{\bf Instance:} Tree $T$, weight function wt: $V(T)
\rightarrow Z^+$, integer $B > 0$.
\item{\bf Question:} Is there an orientation $\vec{T}$ such that
$\mu(\vec{T}) \doteq \sum wt(v) \cdot wt(w) \ge B$, the sum taken
over all pairs $(v,w)$ with a directed path from $v$ to $w$ in
$\vec{T}$?
\end{description}

\smallskip\noindent{\sc\bf Theorem 2.1.}
{\sl WTO is NP-complete.}

\medskip\noindent{\sc\bf Proof.} WTO is obviously in NP, and so
it suffices to show it is NP-hard. For this we reduce PARTITION
[4,p.233] to WTO. Given positive integers $a_1, a_2, \ldots,
a_m$ with even sum $S$, consider the weighted tree $T$ in Fig.~1
and set $B=5(S/2)^2$. Given any $I \subseteq \{1,2,
\ldots, m\}$, consider the orientation $\vec{T}$ obtained by
orienting toward (resp, away from) the vertex with weight $S$
each edge whose other end vertex has weight $a_i$ for $i\in I$
(resp, $i \in \{1,2,\ldots, m\} -I=\overline{I}).$ We then find

$$\mu(\vec{T}) = S\left(\displaystyle\sum_{i\in I} a_i +
\displaystyle\sum_{i\in \overline{I}} a_i\right) +
\displaystyle\sum_{i\in I} a_i \cdot \displaystyle\sum_{i\in
\overline{I}} a_i \le S^2 +
\left(\displaystyle\frac{S}{2}\right)^2 =
5\left(\displaystyle\frac{S}{2}\right)^2$$

\noindent with equality precisely if $\displaystyle\sum_{i \in I}
a_i = \displaystyle\sum_{i\in \overline{I}} a_i
=\displaystyle\frac{S}{2}.$ 
\rule{2mm}{2mm}

\begin{figure}[ht]
\setlength{\unitlength}{.20cm}
\begin{picture}(85,17)\linethickness{.25mm}

\put(33,8){\circle{1.2}}
\put(38,8){\circle{1.2}}
\put(48.2,8){\circle{1.2}}

\put(33,8.7){\line(1,1){6.8}}
\put(38,8.7){\line(1,3){2.2}}
\put(48,8.7){\line(-1,1){7.2}}

\put(40,16){\circle{1.2}}

\put(40,11){\circle{0.1}}
\put(41,11){\circle{0.1}}
\put(42,11){\circle{0.1}}

\put(39,17){\makebox(2,2){S}}
\put(32,4){\makebox(2,2){$a_1$}}
\put(38,4){\makebox(2,2){$a_2$}}
\put(48,4){\makebox(2,2){$a_m$}}
\put(0,0){\makebox(80,3){\bf Figure 1. Tree $T(a_1, \ldots,
a_m)$}}

\end{picture}
\end{figure}

We now wish to develop a key result which will allow us to obtain
a quadratic algorithm for the original orientation problem, as
well as a fully polynomial approximation scheme for WTO. First,
however, we need some terminology and notation. Let $T$ be any
tree with positive integer vertex weights. An {\sl optimal
orientation} $\vec{T}$ is one which maximizes $\mu(\vec{T})$.
Given an orientation $\vec{T}$ and a vertex $w \in V(T)$, define

\medskip\centerline{$\mbox{In}_{\vec{T}}(w)\doteq \{ x \in V(T)~|$
there exists a directed path from $x$ to $w$ in $\vec{T}\}$}

\medskip\centerline{$\mbox{Out}_{\vec{T}}(w)\doteq \{ x \in V(T)~|$
there exists a directed path from $w$ to $x$ in $\vec{T}\}$}

\bigskip In particular, $w \in \mbox{In}_{\vec{T}}(w) \cap
\mbox{Out}_{\vec{T}} (w)$. Define

$$\mbox{In}'_{\vec{T}}(w) \doteq \mbox{In}_{\vec{T}}(w) - \{w\}
\mbox{ and }\, \mbox{Out}'_{\vec{T}}(w) \doteq
\mbox{Out}_{\vec{T}}(w)- \{w\}.$$

\noindent We will usually drop the subscript $\vec{T}$ if it is
clear from the context. Finally, given any subgraph with vertex
set $X \subseteq V(T)$, let $||X|| \doteq \displaystyle\sum_{x \in
X}
wt(x)$ be called the {\sl total weight} of that subgraph. A {\sl
centroid} in a vertex weighted tree is any vertex $c$ whose
removal minimizes the maximum total weight of any component in
$T-c$.

Our goal now is to prove the following.

\bigskip\noindent{\sc\bf Theorem 2.2.} {\sl Let $c$ be a centroid
of $T$. In every optimal orientation $\vec{T}$, we have}
$\mbox{In}_{\vec{T}}(c) \cup \mbox{Out}_{\vec{T}}(c) = V(T)$.

\medskip\noindent Before proving this, we require the following
result.

\bigskip\noindent{\sc\bf Lemma.} {\sl Let $\vec{T}$ be any
optimal orientation of $T$, and let $P$ be any undirected path in
$T$. Then there are no two vertices $x,y$ on $P$ such that the
edges of $P$ incident to $x$ (resp, $y$) are both directed toward
$x$ (resp, away from $y$) in $\vec{T}$.}

\medskip\noindent Briefly, there are no "double reversals" on any
path in $\vec{T}$
(see Fig.~2).

\bigskip\noindent{\sc\bf Proof of the Lemma.} Otherwise, let
$x,y$ be a closest such pair on $P$, so the edges of $P$ between
$x$ and $y$ form a directed path from $y$ to $x$. Let $e_x$
denote the edge at $x$ on the path from $y$ to $x$. Let $X = \{w
\in V(T)|$ there is an undirected
path in $T$ between $w$ and $x$ which does not contain $e_x$\} and
let $X' = X\cap$ In $(x)$.

Let $z$ denote the neighbor of $y$ on $P$ which does not occur on
the directed path from $y$ to $x$ (see Fig.~2). Thinking of $T$
as rooted at $z$, consider the subtree $T_y$ rooted at $y$. Note
that $||\mbox{In}'(y)|| \ge ||\mbox{Out}'(y) \cap T_y ||$ (else
we could reverse the orientations of all edges in $T_y$ to obtain
an orientation better than $\vec{T}$). Reverse all edges in
$\langle X'\rangle$, and consider the gain and loss in
$\mu(\vec{T})$ in doing so, where the loss in $\mu(\vec{T})$ is
the sum of the terms in $\mu(\vec{T})$ which no longer exist
under the new orientation, and the gain in $\mu(\vec{T})$ is
defined analogously. Since $\mbox{Out}'(x) \subset \mbox{Out}'(y)
\cap T_y$, we find

\begin{eqnarray*}
\mbox{Loss in} \,\,\mu(\vec{T}) & \le & ||X'|| \cdot ||
\mbox{Out}'(x)|| < ||X'||\cdot ||\mbox{Out}'(y) \cap T_y|| \\
                                & \le & ||X'||\cdot || 
\mbox{In}'(y) ||\le \mbox{Gain in}\,\,\mu(\vec{T})
\end{eqnarray*}

\noindent and thus we would have a better than optimal
orientation. This proves the Lemma.

\begin{figure}[ht]
\setlength{\unitlength}{.20cm}
\begin{picture}(85,8)\linethickness{.25mm}

\put(13,8){\line(1,0){.25}}
\put(13.5,8){\line(1,0){.25}}
\put(14,8){\line(1,0){.25}}
\put(14.5,8){\line(1,0){.25}}
\put(15,8){\line(1,0){.25}}
\put(15.5,8){\line(1,0){.25}}
\put(16,8){\line(1,0){.25}}
\put(16.5,8){\line(1,0){.25}}

\put(17,8){\circle{1.2}}
\put(18,8){\vector(1,0){4}}
\put(20.7,8){\line(1,0){4}}
\put(25.1,8){\circle{1.2}}
\put(26.2,8){\line(1,0){4}}
\put(34.3,8){\vector(-1,0){4}}
\put(34.5,8){\circle{1.2}}
\put(35.5,8){\line(1,0){.25}}
\put(36,8){\line(1,0){.25}}
\put(36.5,8){\line(1,0){.25}}
\put(37,8){\line(1,0){.25}}
\put(37.5,8){\line(1,0){.25}}
\put(38,8){\line(1,0){.25}}
\put(38.5,8){\line(1,0){.25}}
\put(39,8){\line(1,0){.25}}
\put(39.5,8){\line(1,0){.25}}
\put(40,8){\line(1,0){.25}}
\put(40.5,8){\line(1,0){.25}}
\put(41,8){\line(1,0){.25}}
\put(41.5,8){\line(1,0){.25}}
\put(42,8){\line(1,0){.25}}
\put(42.5,8){\line(1,0){.25}}
\put(43,8){\line(1,0){.25}}
\put(44,8){\circle{1.2}}
\put(45,8){\line(1,0){4.1}}
\put(53,8){\vector(-1,0){4}}
\put(53.2,8){\circle{1.2}}
\put(54.3,8){\vector(1,0){4}}
\put(58.3,8){\line(1,0){4}}
\put(62.5,8){\circle{1.2}}

\put(63.5,8){\line(1,0){.25}}
\put(64,8){\line(1,0){.25}}
\put(64.5,8){\line(1,0){.25}}
\put(65,8){\line(1,0){.25}}
\put(65.5,8){\line(1,0){.25}}
\put(66,8){\line(1,0){.25}}
\put(66.5,8){\line(1,0){.25}}
\put(67,8){\line(1,0){.25}}

\put(10,7){\makebox(2,2){P:}}
\put(24,4){\makebox(2,2){$x$}}
\put(29.3,11){\makebox(2,2){$e_x$}}
\put(52,4){\makebox(2,2){$y$}}
\put(61,4){\makebox(2,2){$z$}}
\put(0,0){\makebox(80,3){\bf Figure 2. Double Reversal on P}}

\end{picture}
\end{figure}

\smallskip\noindent{\sc\bf Proof of Theorem 2.2.} Throughout, think
of $T$ as rooted at centroid $c$. If the theorem fails for some
optimal orientation $\vec{T}$, there must be a vertex $x \not\in
\mbox{In}(c) \cup \mbox{Out}(c)$ adjacent to a vertex $v \in
\mbox{In}'(c) \cup\mbox{Out}'(c)$. (We will call (v,x) a {\sl
dangling edge} at $v$.) We now consider two cases, assuming
$||\mbox{Out}(c)||\ge||\mbox{In}(c)||$ (else reverse the
orientation of all edges of $\vec{T}$).

\medskip\noindent{\sc\bf Case 1.} {\sl There is a dangling edge
$(v,x)$ at $v \in \mbox{In}'(c)$.}

Note that since $c \in \mbox{In}(c) -(T_v \cap \mbox{In}(c))$, we
have

\begin{displaymath}
  \nonumber
||\mbox{Out }(c)||\ge ||\mbox{ In }(c)||>||T_v\cap
 \mbox{ In }(c)||. \end{displaymath}

\noindent Reverse $(v,x)$ and the edges in $T_x$. We find

\begin{eqnarray*}
\mbox{Gain in }\mu (\vec{T})& \ge & ||\mbox{Out}(c)|| \cdot
||\mbox{Out}(x)||\\
& > & ||T_v \cap \mbox{In}(c) || \cdot ||\mbox{Out}(x)||\ge
\mbox{ Loss in } \mu(\vec{T})
\end{eqnarray*}

\noindent contradicting the optimality of $\vec{T}$.

\bigskip\noindent{\sc\bf Case 2.} {\sl There are no dangling
edges at any vertex of $\mbox{In}'(c)$.}

Let $(x,v)$ be a dangling edge at $v \in \mbox{Out}'(c)$. If
$\mbox{outdeg}(c) \ge 2$, then $\vec{T}$ contains a path which
violates the Lemma (with $v,c$ playing the roles of $x,y$ in the
Lemma). Hence we may assume $\mbox{outdeg}(c) = 1$. But then,
since $c$ is a centroid and there are no dangling edges off
vertices in $\mbox{In}'(c)$, and since $\{x\} \cup (T_v \cap
\mbox{Out}(c)) \subset V(T) -\mbox{In}(c)$, we find

\begin{displaymath}
||\mbox{In}(c)|| \ge || V(T) - \mbox{In}(c)|| > ||T_v \cap
\mbox{Out}(c)||. 
\end{displaymath}

\noindent Reverse $(x,v)$ and the edges in $T_x$. We find

$$\mbox{Gain in}\, \mu (\vec{T}) \ge ||\mbox{In}(c)|| \cdot
||\mbox{In}(x)|| > ||T_v \cap \mbox{Out}(c)||\cdot ||\mbox{In}(x)||
\ge \mbox{Loss in}\, \mu(\vec{T})$$
\noindent contradicting the optimality of $\vec{T}$. 
\rule{2mm}{2mm}

\smallskip
It is well known that the centroid of a vertex-weighted tree can
be found in linear time, and that the centroid consists of either
a single vertex or two adjacent vertices [9]. In the latter
case, Theorem 2.2 implies that the optimal orientation may be
represented schematically as in Figure 3, where $c_1$ and $c_2$
denote the adjacent centroids.

\begin{figure}[ht]
\setlength{\unitlength}{.25cm}
\begin{picture}(85,12)\linethickness{.25mm}

\put(19,4){\line(3,4){5}}
\put(19,4){\line(1,0){10}}
\put(29,4){\line(-3,4){5}}

\put(24,11){\circle{1}}

\put(35,4){\line(3,4){5}}
\put(35,4){\line(1,0){10}}
\put(45,4){\line(-3,4){5}}

\put(40,11){\circle{1}}

\put(24.4,11){\vector(1,0){8}}
\put(32,11){\line(1,0){7.7}}
\put(24,5){\vector(0,1){4}}
\put(40,9){\vector(0,-1){4}}

\put(23,12){\makebox(2,2){$c_1$}}
\put(39,12){\makebox(2,2){$c_2$}}

\put(0,0){\makebox(65,3){\bf Figure 3.}}

\end{picture}
\end{figure}


In the former case, we have essentially an instance of PARTITION.
Let $T$ be rooted at $c$, and let $v_1, \ldots, v_k$ denote the
neighbors of $c$. Consider an optimal orientation of $\vec{T}$ of
$T$. By Theorem 2.2, the subtree $T_{v_i}$ must be oriented
entirely toward (respectively, away from) $c$ if and only if the
edge
between $v_i$ and $c$ is oriented toward (respectively, away from)
$c$.
Thus,

$$\mu(\vec{T}) = ||c||\cdot ||T-c|| +
\left(\displaystyle\sum_{i\in I} \big|\big|
T_{v_i}\big|\big|\right) \left(\displaystyle\sum_{i\in
\overline{I}}\big|\big| T_{v_i}\big|\big|\right)
+\displaystyle\sum^k_{i=1}\mu(T^*_{v_i}),$$

\noindent where $I=\{i~|~T_{v_i} \mbox{ is oriented toward } \,c\,
\mbox{ in } \vec{T}\}$ and $T^*_{v_i}$ denotes $T_{v_i}$ oriented
entirely towards (respectively, away from) $c$ if $i\in I$
(respectively,
$i\in\overline{I}$). To maximize $\mu(\vec{T})$, we need to find
a partition $I \cup \overline{I}$ of $\{1, 2, \ldots, k\}$ such
that $\displaystyle\sum_{i\in I} ||T_{v_i}|| \mbox{and}
\displaystyle\sum_{i\in \vec{I}} ||T_{v_i}||$ are as equal as
possible. There is, of course, a well-known dynamic programming
algorithm to find such a partition [4,Section 4.2]. Since
$\displaystyle\sum^{k}_{i=1} ||T_{v_i}|| \le n$, the running time
of this dynamic programming algorithm is $O(kn)=O(n^2)$. Since
all other tasks in the original orientation algorithm (e.g.,
finding the $2$-edge-connected components in $G$, giving these
components strongly-connected orientations, etc.) can easily be
done in $O(|E|)$ time using standard depth-first search
techniques, we see that the original orientation problem can be
solved in $O(n^2)$ time; i.e., quadratic in the original input
size $n$. On the other hand, there is also a well-known fully
polynomial approximation scheme for the above partition problem [7],
which in turn provides a fully polynomial approximation scheme
for the NP-complete problem WTO.

\smallskip\section{\bf Orienting Graphs to Minimize Reachability}

\indent Let $G$ be a connected graph, and let $r(G) = \mbox{min} \,
R(\vec{G})$, the minimum being taken over all orientations of
$G$. We call an orientation $\vec{G}$ a {\sl minimal orientation}
if $R(\vec{G}) = r(G)$. We begin with the following result.

\medskip\noindent{\sc\bf Theorem 3.1.} {\sl Every minimal
orientation of $G$ is acyclic}.

\medskip\noindent{\sc\bf Proof.} Suppose to the contrary there
exists a minimal orientation $\vec{G}$ which is not acyclic.
Since $\vec{G}$ is not acyclic, at least one of the strongly-
connected components of $\vec{G}$, say $C$, is not a single
vertex. Let $E'$ denote a set of edges in $C$ whose reversal
renders $\langle V(C)\rangle$ not strongly connected, and let
$\vec{G}'$ denote $\vec{G}$ with the edges in $E'$ reversed. It
is easy to see every ordered pair of vertices joined by a
directed path in $\vec{G}'$ is joined by a directed path in
$\vec{G}$, while clearly there is a pair of vertices in $V(C)$
which are joined by a directed path in $\vec{G}$ but not in
$\vec{G}'$. Thus, $R(\vec{G}') < R(\vec{G})$, which violates the
minimality of $\vec{G}$. \rule{2mm}{2mm}

Obviously $r(G)\ge|E(G)|$, and it is known that $r(G)=|E(G)|$ if
and only if $G$ is a comparability graph [1,5], or,
equivalently,
if $G$ is transitively orientable [3]. It is known that
comparability graphs can be recognized in polynomial time [3].

Given a connected graph $G$, let $\underline{c} (G)$ (respectively,
$\overline{c}(G)$) denote the minimum number of edges which must
be deleted from (respectively, added to) $G$ to obtain a
comparability
graph. We now apply Theorem 3.1 to establish a connection between
$r(G)$ and $\overline{c}(G)$.

\medskip\noindent{\sc\bf Theorem 3.2.}
$r(G)=|E(G)|+\overline{c}(G)$.

\medskip\noindent{\sc\bf Proof.} Suppose we add $\overline{c}(G)$
edges to $G$ to obtain a comparability graph $G'$, and then
orient $G'$ so that $R(\vec{G'})
=|E(G')|=|E(G)|+\overline{c}(G)$. But, of course $r(G) \le
R(\vec{G}')$, and so $r(G) \le |E(G)|+\overline{c}(G)$.

On the other hand, consider a minimal orientation $\vec{G}$ of
$G$, so that $r(G)=R(\vec{G})$. By Theorem 3.1, $\vec{G}$ is
acyclic. Consider the transitive closure $c\ell (\vec{G})$ of
$\vec{G}$. We then find

\begin{eqnarray*}
r(G)& = & R(\vec{G})=R(c\ell (\vec{G}))\\
& = & |E(G)|+(\mbox{No. of Edges Added to } \vec{G}\,
\mbox{ to obtain } \,c\ell (\vec{G}))\\
& \ge & |E(G)|+\mbox{ (Min. No. of Edges which need to be}\\
& \, & \,\,\,\,\,\mbox{added to G to obtain a transitively
orientable graph) }\\
& = & |E(G)|+\overline{c}(G).
\end{eqnarray*}

\noindent Thus, $r(G)=|E(G)|+\overline{c}(G)$, as asserted. 
\rule{2mm}{2mm}

\medskip
We now turn to the complexity of computing $r(G)$. Consider the
following decision problem.

\medskip\noindent{\bf MINIMUM REACHABILITY ORIENTATION (MRO)}

\begin{description}
\item{\bf Instance:} Graph $G$, integer $k\ge |E(G)|$

\item{\bf Question:} Is $r(G)\le k$?
\end{description}

In a moment we will prove

\medskip\noindent{\sc\bf Theorem 3.3.} {\sl MRO is NP-complete}.

\medskip It follows immediately from Theorems 3.2 and 3.3 that
the following problem is also NP-complete.

\medskip\noindent{\bf COMPARABILITY GRAPH COMPLETION (CGC)}
\begin{description}
\item{\bf Instance:} Graph $G$, integer $k\ge 0$

\item{\bf Question:} Is there a superset $E'$ of $E$ such that
$|E'-E| \le k$ and $G=(V,E')$ is a comparability graph (i.e., is
$\overline{c}(G)\le k)$?
\end{description}

(Previously, it was known only that COMPARABILITY SUBGRAPH (i.e.,
deciding if $\underline{c}(G)\le k$) is NP-complete
[4,p.197].
For the optimization versions of MRO and CGC,
the goal is to compute (or approximate) the minimum $k$ such that
$\langle G,
k\rangle$ is a positive instance of the decision problem.)

\smallskip\noindent{\sc\bf Proof of Theorem 3.3.} Clearly
$\mbox{MRO}\in\mbox{NP}$. To show MRO is NP-hard, we will
reduce NOT-ALL-EQUAL 3SAT [4,p.259].

Let $I$ be an instance of NAE3SAT with $m$ clauses. Construct a
graph $G_I$ as follows. Each variable $x$ will be represented by
an edge $(xT, xF)$ in $G_I$, and orienting this edge towards
$xT$ (respectively, $xF$) will correspond to setting $x$ to $T$
(respectively,
$F$). Each clause $C$ will be represented by a 9-cycle in $G_I$.
The three literals in $C$ will be assigned to three equally
spaced edges $(1T, 1F), (2T, 2F), (3T, 3F)$ on $C's$ 9-cycle, as
shown in Fig.~3. If the literal assigned to the edge $(1T, 1F)$
is $X$ (respectively, $\overline{X}$), add two 2-paths joining $1T$
to
$xF$ and $1F$ to $xT$ (respectively, joining $1T$ to $xT$ and $1F$
to
$xF$) to $G_I$. Make analogous connections for the other two
literals of $C$, as well as for the remaining clauses of $I$, to
complete $G_I$.

\begin{figure}[ht]
\setlength{\unitlength}{.19cm}
\begin{picture}(85,36)\linethickness{.25mm}

\put(15,9){\line(1,0){.5}}
\put(16,9){\line(1,0){.5}}
\put(17,9){\line(1,0){.5}}
\put(20,9){\circle{1}}
\put(20.5,9){\line(1,0){6}}
\put(27,9){\circle{1}}

\put(39.5,21){\line(1,0){7}}
\put(38.7,21.2){\line(-2,3){2}}
\put(36.5,24.5){\line(-1,3){.8}}
\put(36.3,27.3){\line(1,3){1}}
\put(37,30.4){\line(3,2){5.5}}
\put(43.5,33.7){\line(3,-2){5.5}}
\put(49.5,29.6){\line(1,-3){.8}}
\put(50,26.7){\line(-1,-3){1}}
\put(49.1,24){\line(-2,-3){1.7}}

\put(20.5,9.5){\line(3,4){6}}
\put(27.5,18.5){\line(3,4){9}}
\put(36.8,30.2){\circle{1}}
\put(27.5,9.5){\line(1,2){3.5}}
\put(31.5,17.5){\line(1,2){4.5}}
\put(36,27){\circle{1}}

\put(37,26){\makebox(2,2){1F}}
\put(38,29){\makebox(2,2){1T}}

\put(36.7,24){\circle*{1}}

\put(30.5,9){\line(1,0){.5}}
\put(31.5,9){\line(1,0){.5}}
\put(32.5,9){\line(1,0){.5}}
\put(33.5,9){\line(1,0){.5}}
\put(34.5,9){\line(1,0){.5}}
\put(39,9){\circle{1}}
\put(39.5,9){\line(1,0){7}}
\put(47,9){\circle{1}}
\put(41,12){\circle{1}}
\put(45,12){\circle{1}}
\put(39.5,9.5){\line(2,3){1.4}}
\put(41.4,12.5){\line(2,3){5.2}}
\put(46.5,9.5){\line(-2,3){1.4}}
\put(44.4,12.5){\line(-2,3){5.2}}
\put(47,21){\circle{1}}
\put(39,21){\circle{1}}

\put(38.2,22){\makebox(2,2){2T}}
\put(46,22){\makebox(2,2){2F}}
\put(42,18){\makebox(2,2){y}}

\put(27,18){\circle{1}}
\put(31,17){\circle{1}}

\put(49.3,24){\circle*{1}}

\put(51,9){\line(1,0){.5}}
\put(52,9){\line(1,0){.5}}
\put(53,9){\line(1,0){.5}}
\put(54,9){\line(1,0){.5}}
\put(55,9){\line(1,0){.5}}
\put(59,9){\circle{1}}
\put(59.5,9){\line(1,0){6}}
\put(66,9){\circle{1}}

\put(69.5,9){\line(1,0){.5}}
\put(70.5,9){\line(1,0){.5}}
\put(71.5,9){\line(1,0){.5}}

\put(59,9.5){\line(-1,2){3.5}}
\put(54.5,17.5){\line(-1,2){4.5}}
\put(50,27){\circle{1}}
\put(65.5,9.5){\line(-3,4){6}}
\put(58.5,18.5){\line(-3,4){8.2}}

\put(49.5,30){\circle{1}}
\put(47,26){\makebox(2,2){3T}}
\put(46,29){\makebox(2,2){3F}}

\put(34,26){\makebox(1,1){$\overline{x}$}}
\put(51,26){\makebox(1,1){$\overline{z}$}}
\put(55,17){\circle{1}}
\put(59,18){\circle{1}}

\put(43,34){\circle*{1}}

\put(19,6){\makebox(2,2){xT}}
\put(26,6){\makebox(2,2){xF}}
\put(38,6){\makebox(2,2){yT}}
\put(46,6){\makebox(2,2){yF}}
\put(58,6){\makebox(2,2){zT}}
\put(65,6){\makebox(2,2){zF}}

\put(0,0){\makebox(85,3){\bf Figure 4. The Graph $G_I$}}
\put(52,31){\makebox(12,2){$C=\overline{x} +y +\overline{z}$}}

\end{picture}
\end{figure}

Note that however we orient $G_I$, each of the $m$ 9-cycles will
contain a directed 2-path, and thus $r(G_I) \ge |E(G_I)| +m$. In
fact, $r(G_I)=|E(G_I)|+m$ if and only if $I$ has a satisfying
truth assignment in the not-all-equal sense. Indeed, given a
satisfying truth assignment for $I$, we can obtain such an
orientation for $G_I$ as follows: Orient the edges $(xT, xF)$ to
correspond to the truth assignment, and orient all the 6-cycles
in $G_I$ containing  these edges so that none contains a directed
2-path. Note that the edges in each 9-cycle to which a literal
was assigned are now oriented to reflect the truth value of that
literal under the truth assignment. Since each clause contains
both a true and a false literal, it is trivial to complete the
orientation of the 9-cycles so each contains exactly one directed
2-path passing through a darkened vertex on the 9-cycle. 
Conversely, any orientation of $G_I$ with only $|E(G_I)|+m$
reachable pairs
must have exactly one directed 2-path per 9-cycle,
and so corresponds to a satisfying truth assignment.
\rule{2mm}{2mm}

\medskip
We also observe that it remains NP-hard even to approximate
$r(G)$ to within a factor of $1+\epsilon$, for some $\epsilon >0$.
Consider the following optimization problem.

\medskip\noindent{\bf MAX NOT-ALL-EQUAL 3SAT}
\begin{description}
\item{\bf Instance:} Boolean formula in 3CNF
\item{\bf Question:} What is the maximum number of clauses that
can be satisfied (in a not-all-equal sense) by a truth
assignment?
\end{description}

\noindent It was established recently [8,11] that it is NP-hard
to
approximate MAX NAE3SAT to within a factor of 1.013
(that is, finding an
assignment that satisfies, in the not-all-equal sense, 1/1.013 of
the 
maximum possible
number of clauses is NP-hard).
Using this, we can strengthen Theorem 3.3 as follows:

\medskip\noindent{\sc\bf Theorem 3.4.}
{\sl There exists a constant $\epsilon>0$ such that approximating
the optimization versions of MRO or CGC
to within a factor of $1+\epsilon$ is NP-hard.}

\medskip\noindent{\sc\bf Proof Sketch.}
Let $I$ be an instance of NAE3SAT with $m$ clauses.
In the following, ``satisfying'' a clause of $I$
refers to making at least one of its literals true
and at least one false.
Recall that $R(\vec{G_I})$
denotes the number of ordered pairs of vertices $(x,y)$
with a directed path from $x$ to $y$ in an orientation $\vec{G_I}$
of $G_I$.

\medskip
{\noindent\bf Claim 1}:
{\sl Given any assignment satisfying $x$ of the clauses of $I$,
there is an orientation $\vec{G_I}$ of $G_I$
such that $R(\vec{G_I}) = |E(G_I)| + 3m - 2x$.}

\medskip
Claim 1 follows from the proof of Theorem 3.3.

\medskip
{\noindent\bf Claim 2}: 
{\sl Given any orientation $\vec{G_I}$,
there is an assignment satisfying at least
$\frac{1}{2}(|E(G_I)| + 3m - R(\vec{G_I}))$
clauses of $I$.}
\medskip
Claim 2 holds because if any clause-subgraph (see Fig.~4)
in $G_I$ has more than one directed 2-path, then it has at least
three.
Thus, any orientation can be converted into an equally good
orientation corresponding to a truth assignment (where the only
directed
2-paths are 2-paths through the darkened vertices on the 9-cycles).
By simple algebra, the claimed bound holds for this assignment.

\medskip
Suppose one could find an orientation $\vec{G_I}$ of $G_I$
with $R(G_I) \le (1+\epsilon)r(G_I)$.
By Claim 2, this would yield an assignment satisfying at least 
\begin{equation}
  \frac{1}{2}\big[|E(G_I)| + 3m - (1+\epsilon)r(G_I)\big]
\label{eqn1}
\end{equation}
clauses of $I$.
Let $\mbox{max}(I)$ denote the maximum number of
clauses in $I$ which can be simultaneously satisfied in a 
not-all-equal sense.  
By Claim 1, $r(G_I) \le |E(G_I)| + 3m - 2\max(I)$.
Thus expression (\ref{eqn1}) is at least $ \displaystyle
\frac{1}{2} \left[|E(G_I)| + 3m -
(1+\epsilon)(|E(G_I)| + 3m - 2\max(I))\right]$, which equals
\begin{equation}
  \max(I)-\frac{\epsilon}{2}\big[|E(G_I)| + 3m -
2\max(I)\big].\label{eqn2}
\end{equation}
Using $|E(G_I)| \le 24m$ and $m \le 2\max(I)$ (assuming without
loss of
generality that every clause has at least two literals, a random
assignment
satisfies at least half the clauses on average),
expression (\ref{eqn2}) is at least
$ \max(I) - \displaystyle\frac{\epsilon}{2}\big[48\max(I) +
6\max(I) 
- 2\max(I)\big]
 = (1 - 26\epsilon)\max(I)$.
From this the claimed hardness of approximating MRO follows,
with $\epsilon > 0.00049$.

A similar argument, using Theorem 3.2 and omitting the $|E(G_I)|$
term
in expression (\ref{eqn1}) (and subsequent expressions),
establishes the claimed hardness of approximating
CGC,
with $\epsilon > 0.0064$.
\rule{2mm}{2mm}

\noindent\section{\bf Concluding Remarks}

It would be interesting to determine the algorithmic complexity
of completing the orientation of a partially oriented graph to
maximize reachability. Thus far we have made little progress on
this problem. The analogous completion problem to minimize
reachability is, of course, NP-hard.

On the other hand, it is very easy to characterize partial
orientations which can be completed into strong orientations. An
edge-cut $(X,\overline{X})$ in a partially oriented graph is
called {\sl one-way} if all the edges in the cut are already
oriented, and all are oriented from $X$ to $\overline{X}$ or all
from $\overline{X}$ to $X$. We have

\smallskip\noindent{\sc\bf Theorem 4.1.}
{\sl A partial orientation of a
graph $G$ can be completed into a strong orientation if and only
if $G$ is 2-edge-connected and there are no one-way edge cuts in
the partial orientation.}


\section{\bf References} 
 
\begin{description} 
\setlength{\itemsep}{0cm}\item{[1]} C. Berge, {\sl Graphs and
Hypergraphs,} North-Holland, Amsterdam, 1973.
\item{[2]} J.A. Bondy and U.S.R. Murty, {\sl Graph Theory with
Applications,} North-Holland, Amsterdam, 1976.
\item{[3]} S. Even, A. Pnueli, and A. Lempel, "Permutation graphs
and transitive graphs," {\sl J. Assoc. Comp. Mach.} {\bf 19}
(1972), 400-410.
\item{[4]} M.R. Garey and D.S. Johnson, {\sl Computers and
Intractability: A Guide to NP-Completeness,} W.H. Freeman, San
Francisco, 1979.
\item{[5]} P.C. Gilmore and A.J. Hoffman, "A characterization of
comparability graphs and of interval graphs," {\sl Can. J. Math.}
{\bf 16} (1964), 539-548.
\item{[6]} M.C. Golumbic, H. Kaplan and R. Shamir, "Graph sandwich
problems," {\sl J. Algorithms} {\bf 19} (1995), 449-473.
\item{[7]} O.H. Ibarra and C.E. Kim, "Fast approximation algorithms
for the knapsack and sum of subset problems," {\sl J. Assoc. Comp.
Mach.} {\bf 22} (1975), 463-468.
\item{[8]} V. Kann, J. Lagergren, and A. Panconesi,
"Approximability of maximum splitting of $k$-sets and some other
APX-complete problems," {\sl Tech. Report TRITA-NA-P9509,} Dept. of
Numer. Analysis and Comp. Sci., Royal Inst. of Technology,
Stockholm.
\item{[9]} O. Kariv and S.L. Hakimi, "An algorithmic approach to
network location problems, II: the p-medians,"{\sl SIAM J. Appl.
Math.}{\bf 37}(1979), 539-560.
\item{[10]} P. Dell'Olmo, M. Grazia Speranza and Z. Tuza,
"Comparability graph augmentation for some multiprocessor
scheduling problems," {\sl Discrete Appl. Math.} {\bf 72} (1997),
71-84.
\item{[11]} C.H. Papadimitriou and M. Yannakakis, "Optimization,
approximation, and complexity classes," {\sl J. Comp. Systems
Sci.,} {\bf 43} (1991), 425-440.
\item{[12]} H.E. Robbins, "A theorem on graphs with an application
to a problem of traffic control," {\sl Amer. Math. Monthly} {\bf
46} (1939), 281-283.

\end{description} 
 
\end{document}